# Simulation of quantum gates by postselection of interference experiments in multiport beam-splitter (BS) configurations

Y. Ben-Aryeh

*Department of Physics, Technion-Israel Institute of Technology, Haifa 32000, Israel*
*(phr65yb@physics.technion.ac.il)*

Using multiport beam-splitter (BS) configurations eight input operators $\hat{a}_0, \hat{a}_1, \hat{a}_2 \ldots \hat{a}_7$ are mixed by 12 BS's to produce the output operators $\hat{b}_0, \hat{b}_1, \hat{b}_2 \ldots \hat{b}_7$. A single photon entering into, or exiting from, one port of a BS is considered as the $|0\rangle$ state while that in the second port of the BS is considered as the $|1\rangle$ state. Two single photons are inserted into two of the BS's simulating two input qubit-states by operators $\hat{a}_0, \hat{a}_1, \hat{a}_2, \hat{a}_3$. In order to simulate the two-qubit gates we exclude by postselection all the cases in which one or two photons exit through output ports $\hat{b}_4, \hat{b}_5, \hat{b}_6, \hat{b}_7$ so that only 25% of the experiments are taken into account. The method simulates explicitly the *CNOT* and *SWAP* gates but similar methods can be used for other gates.

## 1. INTRODUCTION

Various quantum computation processes can be simulated by the use of optics' interferometry [1-13]. It has been shown that by using multiport beam-splitter (BS) configurarions one can realize any unitary transformation operating on *single photons* [2-4]. In particular such configurations can realize the quantum discrete Fourier transform [2-6] which is an important component in quantum computation algorithms [14,15]. The main difficulty in realizing quantum computation processes with photons is, however, the optical implementation of 'controlled' interactions between photonic qubits. Controlled interactions between separate single photons would require strong non-linearities which are beyond the use of the present technology. Recently, it has been shown that the desired interactions between photonic qubits can be realized by postselection [16-18]. Various devices for implementing quantum gates by postselection have been analyzed in many works [19-43] by following this idea [16-18]. In the present work we study the use of linear optics with two photons in multiport BS configurations for simulating quantum gates by the use of postselection.

A quantum bit (qubit) is a two-level quantum system described by a two-dimensional complex Hilbert space. In this space the computational basis is represented by a pair of normalized and mutually orthogonal quantum states denoted as $|0\rangle$ and $|1\rangle$. There are various ways to implement qubits [14,15]. In the present study of simulating quantum gates with multiport BS configurations with postselection one single photon is inserted in an input BS. Inserting a single photon in one input port of a beam splitter is considered as the $|0\rangle$ state while if the single photon is inserted in the second input port it is considered as the $|1\rangle$ state. One should take into account that in any two-level system one can use a specific definition of the $|0\rangle$ and $|1\rangle$ states. From the superposition principle, any state of the qubit may be written as

$$|\psi\rangle = \alpha|0\rangle + \beta|1\rangle \quad , \tag{1}$$

where the amplitudes $\alpha$ and $\beta$ are complex numbers constrained by the normalization condition $|\alpha|^2 + |\beta|^2 = 1$. Such superposition can be also inserted in the two input ports of a BS.

It has been shown in previous studies [2-6] that by using one-photon linear transformations various effects in quantum computation can be implemented. We are interested here, however, in *controlled operations*. The prototypical control operation is the cotrolled-NOT (*CNOT*). This is a quantum gate with two input qubits, known as the *control qubit* and the *target qubit*, respectively. In terms of computational basis, the action of the CNOT is given by $|c\rangle|t\rangle \rightarrow |c\rangle|t \oplus c\rangle$; that is if the control bit is set to $|1\rangle$ then the target bit is flipped, otherwise the target qubit is left alone. Quantum computations [14,15] are based on two-qubit gates including in particular the *CNOT* gate.

For realizing quantum computation processes one needs to use two-qubit quantum gates and optics seems to be a prominent candidate for achieving two-qubit quantum gates. Unfortunately, such gates including the basic *CNOT* gate [14,15] are very difficult to realize experimentally. This problem can be explained as follows.
An input two-qubit state can be written as

$$|\psi\rangle_{in} = \{\alpha|0\rangle_A|0\rangle_B + \beta|0\rangle_A|1\rangle_B + \gamma|1\rangle_A|0\rangle_B + \delta|1\rangle_A|1\rangle_B\} \quad , \tag{2}$$

where the subscripts A and B refer to two qubits. $|\psi\rangle_{in}$ is given by a superposition of two-qubit states with corresponding amplitudes $\alpha, \beta, \gamma$ and $\delta$. The *CNOT* gate is defined as leading to the output state

$$|\psi\rangle_{out} = \{\alpha|0\rangle_A|0\rangle_B + \beta|0\rangle_A|1\rangle_B + \gamma|1\rangle_A|1\rangle_B + \delta|1\rangle_A|0\rangle_B\} \quad . \tag{3}$$

The first bit (A) acts as a control and its value is unchanged on the output. The second (target) bit (B) is flipped ($|0\rangle \rightarrow |1\rangle$; $|1\rangle \rightarrow |0\rangle$) if and only if the first bit is set to one. Such gate is quite difficult to implement since the state of the control qubit should affect the second target qubit and this requires strong interactions between single photons. Such interactions need high nonlinearities well beyond what is available experimentally. Similar problems arise for other two-qubit gates.

For using matrix representations of quantum gates the qubits $|0\rangle$ and $|1\rangle$ are described by the following column vectors

$$|0\rangle \equiv \begin{pmatrix} 1 \\ 0 \end{pmatrix} \quad ; \quad |1\rangle \equiv \begin{pmatrix} 0 \\ 1 \end{pmatrix} \quad . \tag{4}$$

The transformations operating on these single qubit column vectors can be given by multiplying them by unitary matrices of dimension $2\times 2$:

$$I = \begin{pmatrix} 1 & 0 \\ 0 & 1 \end{pmatrix}, \quad \sigma_1 = \begin{pmatrix} 0 & 1 \\ 1 & 0 \end{pmatrix}, \quad \sigma_2 = \begin{pmatrix} 0 & -i \\ i & 0 \end{pmatrix}, \quad \sigma_3 = \begin{pmatrix} 1 & 0 \\ 0 & -1 \end{pmatrix}, \quad (5)$$

where $I$ is the two-dimensional unit matrix, and $\sigma_1, \sigma_2, \sigma_3$ are the Pauli spin matrices.

The two-qubit state can be given by four dimensional column vectors:

$$|00\rangle \equiv \begin{pmatrix} 1 \\ 0 \end{pmatrix} \otimes \begin{pmatrix} 1 \\ 0 \end{pmatrix} \equiv \begin{pmatrix} 1 \\ 0 \\ 0 \\ 0 \end{pmatrix} \; ; \quad |01\rangle \equiv \begin{pmatrix} 1 \\ 0 \end{pmatrix} \otimes \begin{pmatrix} 0 \\ 1 \end{pmatrix} \equiv \begin{pmatrix} 0 \\ 1 \\ 0 \\ 0 \end{pmatrix}$$

$$|10\rangle \equiv \begin{pmatrix} 0 \\ 1 \end{pmatrix} \otimes \begin{pmatrix} 1 \\ 0 \end{pmatrix} \equiv \begin{pmatrix} 0 \\ 0 \\ 1 \\ 0 \end{pmatrix} \; ; \quad |11\rangle \equiv \begin{pmatrix} 0 \\ 1 \end{pmatrix} \otimes \begin{pmatrix} 0 \\ 1 \end{pmatrix} \equiv \begin{pmatrix} 0 \\ 0 \\ 0 \\ 1 \end{pmatrix} \quad (6)$$

where in the ket states on the left handside of these equations the first and second number denote the state of the first and second qubit, respectively. The sign $\otimes$ represents tensor product where the two-qubit states can be described by tensor products of the first and second qubit column vectors.

The $CNOT$ gate operates on the two-qubit states as

$$CNOT\,|00\rangle = |00\rangle,\; CNOT\,|01\rangle = |01\rangle,\; CNOT\,|10\rangle = |11\rangle,\; CNOT\,|11\rangle = |10\rangle. \quad (7)$$

Any two-qubit gate operating on the two qubit states can be given by a unitary matrix $U_2$ of dimension $4\times 4$ multiplying the 4 dimensional vectors describing the two-qubit states by (6). The $CNOT$ gate is operating on the four dimensional vectors by the unitary matrix

$$CNOT = \begin{pmatrix} 1 & 0 & 0 & 0 \\ 0 & 1 & 0 & 0 \\ 0 & 0 & 0 & 1 \\ 0 & 0 & 1 & 0 \end{pmatrix} \quad . \tag{8}$$

All other two-qubit gates are described by the corresponding unitary $4 \times 4$ matrices.

In another example the *SWAP* gate operating on the two qubit states is given by

$$SWAP |00\rangle = |00\rangle, \; SWAP |01\rangle = |10\rangle, \; SWAP |10\rangle = |01\rangle, \; SWAP |11\rangle = |11\rangle \quad . \tag{9}$$

By this gate the states $|01\rangle$ and $|10\rangle$ are exchanged ($|01\rangle \to |10\rangle, |10\rangle \to |01\rangle$) and nothing happens to the states $|00\rangle$ and $|11\rangle$. The SWAP gate is operating on the four dimensional vectors of (6) by the unitary matrix

$$SWAP = \begin{pmatrix} 1 & 0 & 0 & 0 \\ 0 & 0 & 1 & 0 \\ 0 & 1 & 0 & 0 \\ 0 & 0 & 0 & 1 \end{pmatrix} \quad . \tag{10}$$

In the present work we analyze the possibility to use interference experiments with postselection for simulating the *CNOT* and the *SWAP* gates. By using similar methods we can, however, simulate also other quantum gates.

The unitary transformations obtained by multiport beam splitter (BS) configurations [2-6] are operating only on single photons represented by the two dimensional vectors $|0\rangle$ and $|1\rangle$ and *they cannot realize directly* the quantum gates based on controlled operations of two qubits where the unitary $4 \times 4$ matrices are operating on the four dimensional vectors of (6). It is suggested in the present work *to add postselection to certain interference experiments* which will simulate quantum gates. The equalities and differences between the present method for simulating quantum gates and the ordinary implementation of quantum gates will be analyzed.

The idea and the organization of the present work is as follows:

By using two input BS's we can simulate each of the input two-qubit states $|00\rangle, |01\rangle, |10\rangle, |11\rangle$ by two single photons according to the input port of each BS into which the single photons enters. The output two-qubit states are defined according to the output port of each BS into which the corresponding single photon is exiting. We will study in Section 2 *non-unitary* matrices of dimension $4 \times 4$ which will transform *mathematically* these input states to output states and which will simulate the quantum gate *CNOT* or *SWAP*. Since we are supposed by quantum mechanics to use only unitary transformations, we will extend the $4 \times 4$ *non-unitary* matrices into $8 \times 8$ unitary matrix where the $4 \times 4$ *non-unitary* is a part of the extended matrix. Then we show in Section 3 how to simulate by multiport BS's configurations the extended $8 \times 8$ unitary matrix and how to simulate the *CNOT* or *SWAP* gate, by postselection. In Section 4 we present a discussion and summary of our results.

## 2. TWO-QUBITS' QUANTUM GATES SIMULATED MATHEMATICALLY BY NON-UNITARY TRANSFORMATIONS OPERATING ON TWO-PHOTONS

The qubits defined by (4) can be realized by a single photon entering into an input port of a BS. The state $\begin{pmatrix} 1 \\ 0 \end{pmatrix}$ will represent a single photon entering the first input port of the BS and zero photons are entering into the second input port ; vice versa, the state $\begin{pmatrix} 0 \\ 1 \end{pmatrix}$ is represented by a single photon entering into the second input port and zero photons into the first input port. The quantum gates are obtained by applying unitary transformations which include 'controlled' operations on the four dimensional vectors of (6). For applying interference experiments we will use a different basis of states for the four dimensional vectors which 'simulates' the two-qubit states by a superposition of two-photons' states and which can be applied by the use of multiport BS's configurations:

$$|00\rangle \equiv \begin{pmatrix}1\\0\end{pmatrix}_A \begin{pmatrix}1\\0\end{pmatrix}_B \rightarrow (\hat{a}_1^\dagger \hat{a}_3^\dagger)|0\rangle = \begin{pmatrix}1\\0\\1\\0\end{pmatrix} \quad ; \quad |01\rangle \equiv \begin{pmatrix}1\\0\end{pmatrix}_A \begin{pmatrix}0\\1\end{pmatrix}_B \rightarrow (\hat{a}_1^\dagger \hat{a}_4^\dagger)|0\rangle = \begin{pmatrix}1\\0\\0\\1\end{pmatrix};$$

$$|10\rangle \equiv \begin{pmatrix}0\\1\end{pmatrix}_A \begin{pmatrix}1\\0\end{pmatrix}_B \rightarrow (\hat{a}_2^\dagger \hat{a}_3^\dagger)|0\rangle = \begin{pmatrix}0\\1\\1\\0\end{pmatrix}; \quad |11\rangle \equiv \begin{pmatrix}0\\1\end{pmatrix}_A \begin{pmatrix}0\\1\end{pmatrix}_B \rightarrow (\hat{a}_2^\dagger \hat{a}_4^\dagger)|0\rangle = \begin{pmatrix}0\\1\\0\\1\end{pmatrix}$$

.(11)

In (11) we have two BS's denoted by the subscripts A and B, where in each BS one photon is inserted. The state $|00\rangle$ corresponds to two single photons where each of the single photon is inserted into the first input port of the corresponding BS. The state $|01\rangle$ corresponds to the first single photon entering the first input port of the first BS while the second single photon is entering into the second input port of the second BS. Vice versa, the state simulating $|10\rangle$ corresponds to the first single photon entering the second input port of the first BS while the second single photon is entering the first input port of the second BS. The state $|11\rangle$ corresponds to two single photons where each of the single photons is entering into the second input port of the corresponding BS.

One should notice the fundamental difference between the four-dimensional vectors of (6) for implementing 'controlled' quantum gates and those given in (11) for 'simulating' 'controlled' interference experiments. While in (6) (represented also by the left handside of Eq.(11)) the quantum basis of states is given by a <u>direct product</u> of the states of the two qubits, on the right handside of (11) the basis of states for interference experiments is represented by <u>a</u>

multiplication of two photon states. The location of the ones' in these states represent the ports into which the photons are entering into , or exiting out, the BS's. The location of the zeros' represent the ports which are "empty" ,i.e., without any photons. We will use in the present study multiport BS's configurations for transforming the four dimensional vectors of (11).

One should notice also that we define here states of two single-photons and therefore we use a different definition from that used for single photons. In using orthonormality conditions one should take into account that states related to different BS's are orthogonal,

By using interference experiments (without any losses) with multiport BSs' configurations the two-qubits input quantum states given by (11) are transformed into two-qubits output states. It is quite easy to find that such unitary transformations implemented by interference experiments cannot simulate the controlled operations of quantum gates (e.g. *CNOT*, *SWAP*) implemented by unitary transformations operating on the four dimensional basis of (6) We find ,however, the *interesting point* that *non-unitary* $4 \times 4$ matrices operating on the four dimensional vectors of (11) can *simulate mathematically* the quantum gates *CNOT* and *SWAP*. We will analyze in the present section such *non-unitary* transformations. In the next section we will show how to obtain *physically* such non-unitary matrices by postselection.

Simulation of the *CNOT* gate in interference experiments up to relative phase corresponds to the *non-unitary* matrix transformation

$$A_{CNOT} = \frac{1}{2} \begin{pmatrix} 1 & -1 & 1 & 1 \\ -1 & 1 & 1 & 1 \\ -1 & 1 & -1 & 1 \\ 1 & -1 & -1 & 1 \end{pmatrix} ,  \qquad (12)$$

operating on the four dimensional states of (11), but notice that only two ports are occupied in the four dimensionl vectors of (11). The operation of $A_{CNOT}$ on the two-photon states of (11) are given as:

$$A_{CNOT}\begin{pmatrix}1\\0\\1\\0\end{pmatrix}=\begin{pmatrix}1\\0\\-1\\0\end{pmatrix} \quad ; \quad A_{CNOT}\begin{pmatrix}1\\0\\0\\1\end{pmatrix}=\begin{pmatrix}1\\0\\0\\1\end{pmatrix} \quad ;$$

$$A_{CNOT}\begin{pmatrix}0\\1\\1\\0\end{pmatrix}=\begin{pmatrix}0\\1\\0\\-1\end{pmatrix} \quad ; \quad A_{CNOT}\begin{pmatrix}0\\1\\0\\1\end{pmatrix}=\begin{pmatrix}0\\1\\1\\0\end{pmatrix} \quad .$$

(13)

In the 4 dimensional vectors of (13) the location of the ones represent the location of the input ports or output ports into which the two photons are entering into, or exiting out, respectively. The zeros in these vectors represent the ports which are empty ,i.e., without any photons. Using the correspondence between the four dimensional states of (13) and the two-qubit states of (6) according to the correspondence of (11) we find that the non-unitary matrix simulates mathematically, up to relative phase, the CNOT gate. Quantum gates up to relative phases (or signs) have been referred as 'mapping operations' (see [14] pp.320,183) and can be used in quantum computation. We find ,however, that the simulation of the $C_{NOT}$ gate will simulate the output of the ordinary *CNOT* gate if we can invert the relative phase for the output states by $\pi$ for the cases in which a single photon is entering into the first input port of the second BS.

The *SWAP* gate corresponds to the *non-unitary* matrix transformation:

$$A_{SWAP}=\frac{1}{2}\begin{pmatrix}1 & -1 & 1 & 1\\-1 & 1 & 1 & 1\\1 & -1 & 1 & 1\\-1 & 1 & 1 & 1\end{pmatrix} \quad ,$$

(14)

operating on the four dimensional states of (11), but notice that only two ports are occupied in these states. The operation of $A_{SWAP}$ on the two-photon states of (11) are given as:

$$A_{SWAP}\begin{pmatrix}1\\0\\1\\0\end{pmatrix}=\begin{pmatrix}1\\0\\1\\0\end{pmatrix} \quad ; \quad A_{SWAP}\begin{pmatrix}1\\0\\0\\1\end{pmatrix}=\begin{pmatrix}1\\0\\1\\0\end{pmatrix} \quad ;$$

$$A_{SWAP}\begin{pmatrix}0\\1\\1\\0\end{pmatrix}=\begin{pmatrix}0\\1\\0\\1\end{pmatrix} \quad ; \quad A_{SWAP}\begin{pmatrix}0\\1\\0\\1\end{pmatrix}=\begin{pmatrix}0\\1\\0\\1\end{pmatrix} \quad .$$
(15)

Here again the location of the ones represent the location of the input ports or output ports into which the two photons are entering into, or exiting out, respectively. The zeros in these vectors represent the ports which are "empty". Using the correspondence between the four dimensional vectors vectors of (15) and those of (6) according to (11) for these two sets of two-qubit states we find that the non-unitary matrix $A_{SWAP}$ *simulates mathematically* the SWAP gate.

For mixing quantum states by multiport BS configurations it is efficient to describe the transformation as operating on operators. Such operators are given in pairs $\hat{a}_0$, $\hat{a}_1$ ($\hat{b}_0$, $\hat{b}_1$) belonging to the first input (output) BS, $\hat{a}_2$, $\hat{a}_3$ ($\hat{b}_2$, $\hat{b}_3$) belonging to the second input (output) BS etc. . The operation of $A_{CNOT}$ and $A_{SWAP}$ operating on the four dimensional states of (11) given, respectively, in (13) and (15) can be presented as operation on the four dimensional operatoric vectors given by the correspondence:

$$\begin{pmatrix}1\\0\\1\\0\end{pmatrix} \to \begin{pmatrix}\hat{a}_0^{\dagger}\\0\\a_2^{\dagger}\\0\end{pmatrix} \quad ; \quad \begin{pmatrix}1\\0\\0\\1\end{pmatrix} \to \begin{pmatrix}\hat{a}_0^{\dagger}\\0\\0\\\hat{a}_3^{\dagger}\end{pmatrix} \quad ; \quad \begin{pmatrix}0\\1\\1\\0\end{pmatrix} \to \begin{pmatrix}0\\a_1^{\dagger}\\a_2^{\dagger}\\0\end{pmatrix} \quad ; \quad \begin{pmatrix}0\\1\\0\\1\end{pmatrix} \to \begin{pmatrix}0\\a_1^{\dagger}\\0\\a_3^{\dagger}\end{pmatrix}. \quad (16)$$

One should notice that by operating with the four dimensional operatoric vectors given on the right handside of (16) on the vacuum we get the photon states given on the left side of this equation. The operation of the matrices $A_{CNOT}$ of (12) and $A_{SWAP}$ of (14) can be described as operating on the four dimensional vectors of (16) leading to the correct states when the creation operators are operating on the vacuum. More generally the operation of $A_{CNOT}$ and $A_{SWAP}$ on the creation operators are given ,respectively, by

$$A_{CNOT}\begin{pmatrix}\hat{a}_0^{\dagger}\\\hat{a}_1^{\dagger}\\\hat{a}_2^{\dagger}\\\hat{a}_3^{\dagger}\end{pmatrix} = \begin{pmatrix}\hat{b}_0^{\dagger}\\\hat{b}_1^{\dagger}\\\hat{b}_2^{\dagger}\\\hat{b}_3^{\dagger}\end{pmatrix} \quad ; \quad A_{SWAP}\begin{pmatrix}\hat{a}_0^{\dagger}\\\hat{a}_1^{\dagger}\\\hat{a}_2^{\dagger}\\\hat{a}_3^{\dagger}\end{pmatrix} = \begin{pmatrix}\hat{b}_0^{\dagger}\\\hat{b}_1^{\dagger}\\\hat{b}_2^{\dagger}\\\hat{b}_3^{\dagger}\end{pmatrix} \quad , \quad (17)$$

where $\hat{b}_0^{\dagger}, \hat{b}_1^{\dagger}, \hat{b}_2^{\dagger}, \hat{b}_3^{\dagger}$ are the output operators but since only two ports are occupied by photons the four dimensional vectors of (17) includes only two creation operators and two zeros as those given by (16). Since the transformation matrices $A_{CNOT}$ of (12) and for $A_{SWAP}$ of (14) include only real elements the same transformation matrices will operate also on the annihilation operatrs. Therefore for the simplicity of notation we will relate from now on the transformation as operating on annihilation operators. In the present Section it has been shown how to simulate quantum gates with non-unitary transformations operating on the basis of four dimensional states given by (11). In the next Section it is shown how to obtain such non-unitary transformation from extended matrix of transformation using postselection.

## 3. TWO-QUBITS' QUANTUM GATES SIMULATED BY INTERFERENCE USING MULTIPORT CONFIGURATIONS WITH POSTSELECTION

In order to simulate the quantum gates by extended unitary matrix with postselection we assume to have initially 8 input operators $\hat{a}_0$, $\hat{a}_1$, $\hat{a}_2$ ... $\hat{a}_7$ which are mixed by the multiport configurations to produce the output operators $\hat{b}_0$, $\hat{b}_1$, $\hat{b}_2$ ... $\hat{b}_7$ which will be related to the input operators by a unitary transformation. One should notice that the eight operators are given in pairs $\hat{a}_0$, $\hat{a}_1$ ($\hat{b}_0$, $\hat{b}_1$) belonging to the first input (output) BS, $\hat{a}_2$, $\hat{a}_3$ ($\hat{b}_2$, $\hat{b}_3$) belonging to the second input (output) BS etc. . It has been shown in [2] that any unitary transformation can be implemented by interference with multiport configurations. We will use, however a special method for realizing such transformation.

The method is based on performing the unitary transformation in 3 stages where in each stage the 8'th input operators are mixed by 4 BS's into 8'th output operators including in total 12 BS's. This method has been used previously [6] to implement 3 qubits discrete Fourier transform (DFT) . Here we use such $8 \times 8$ unitary transformations to simulate two-qubit controlled gates for two single-photons by postselection.

While in the general case the above transformations are quite complicated, for the purpose of simulating by postselection the *CNOT* and *SWAP* gates we can simplify the analysis by assuming that each BS produces the simple transformation

$$\begin{pmatrix} \hat{b}_{1out} \\ \hat{b}_{2out} \end{pmatrix} = \begin{pmatrix} \cos\theta & \sin\theta \\ -\sin\theta & \cos\theta \end{pmatrix} \begin{pmatrix} \hat{a}_{1in} \\ \hat{a}_{2in} \end{pmatrix} \quad . \tag{18}$$

The 12 BS's used in this analysis are characterized by the corresponding angles $\theta_1, \theta_2, \theta_3 \cdots \theta_{12}$. We might generalize the BS transformation (18) for simulating other quantum two-qubit gates.

The 8'th operators obtained after the two first stages of 8 BS's can be written as:

$$\begin{aligned}
\hat{c}_0 &= \{\cos\theta_5 [\hat{a}_0 \cos\theta_1 + \hat{a}_4 \sin\theta_1] + \sin\theta_5 [\hat{a}_2 \cos\theta_3 + \hat{a}_6 \sin\theta_3]\} ; \\
\hat{c}_1 &= \{\cos\theta_6 [\hat{a}_1 \cos\theta_2 + \hat{a}_5 \sin\theta_2] + \sin\theta_6 [\hat{a}_3 \cos\theta_4 + \hat{a}_7 \sin\theta_4]\} ; \\
\hat{c}_2 &= \{\cos\theta_7 [\hat{a}_4 \cos\theta_1 - \hat{a}_0 \sin\theta_1] + \sin\theta_7 [\hat{a}_6 \cos\theta_3 - \hat{a}_2 \sin\theta_3]\} ; \\
\hat{c}_3 &= \{\cos\theta_8 [\hat{a}_5 \cos\theta_2 - \hat{a}_1 \sin\theta_2] + \sin\theta_8 [\hat{a}_7 \cos\theta_4 - \hat{a}_3 \sin\theta_4]\} ; \\
\hat{c}_4 &= \{\cos\theta_5 [\hat{a}_2 \cos\theta_3 + \hat{a}_6 \sin\theta_3] - \sin\theta_5 [\hat{a}_0 \cos\theta_1 + \hat{a}_4 \sin\theta_1]\} ; \quad (19) \\
\hat{c}_5 &= \{\cos\theta_6 [\hat{a}_3 \cos\theta_4 + \hat{a}_7 \sin\theta_4] - \sin\theta_6 [\hat{a}_1 \cos\theta_2 + \hat{a}_5 \sin\theta_2]\} ; \\
\hat{c}_6 &= \{\cos\theta_7 [\hat{a}_6 \cos\theta_3 - \hat{a}_2 \sin\theta_3] - \sin\theta_7 [\hat{a}_4 \cos\theta_1 - \hat{a}_0 \sin\theta_1]\} ; \\
\hat{c}_7 &= \{\cos\theta_8 [\hat{a}_7 \cos\theta_4 - \hat{a}_3 \sin\theta_4] - \sin\theta_8 [\hat{a}_5 \cos\theta_2 - \hat{a}_1 \sin\theta_2]\} .
\end{aligned}$$

One should notice that unitary transformations given in the square brackets in the first 4'th lines of (20) represent the first stage of using 4 BS's transformations with corresponding angles $\theta_1, \theta_2, \theta_3, \theta_4$. A second stage of transformations is represented in the 8'th curled brackets of (20) by using additional 4 BS's with corresponding angles $\theta_5, \theta_6, \theta_7, \theta_8$. By using the CR $[\hat{a}_i, \hat{a}_j] = \delta_{i,j}$ it is quite easy to verify that also $\hat{c}_i$ $(i = 0, 1, \cdots, 7)$ satisfy the CR $[\hat{c}_i, \hat{c}_j] = \delta_{i,j}$. The arrangement of the BS's which will produce the transformation (19) is straightforward and can be implemented by following similar arrangement to that presented for the discrete Fourier transform (see [6], Fig. 2).

The third stage of transformations is given as:

$$\hat{b}_0 = \hat{c}_0 \cos\theta_9 + \hat{c}_1 \sin\theta_9 \quad;$$
$$\hat{b}_1 = \hat{c}_4 \cos\theta_{10} + \hat{c}_5 \sin\theta_{10} \quad;$$
$$\hat{b}_2 = \hat{c}_2 \cos\theta_{11} + \hat{c}_3 \sin\theta_{11} \quad;$$
$$\hat{b}_3 = \hat{c}_6 \cos\theta_{12} + \hat{c}_7 \sin\theta_{12} \quad;$$
$$\hat{b}_5 = \hat{c}_1 \cos\theta_9 - \hat{c}_0 \sin\theta_9 \quad;\qquad(20)$$
$$\hat{b}_6 = \hat{c}_5 \cos\theta_{10} - \hat{c}_4 \sin\theta_{10} \quad;$$
$$\hat{b}_7 = \hat{c}_3 \cos\theta_{11} - \hat{c}_2 \sin\theta_{11} \quad;$$
$$\hat{b}_8 = \hat{c}_7 \cos\theta_{12} - \hat{c}_6 \sin\theta_{12} \quad,$$

where it is easy to verify that also $\hat{b}_i$ $(i = 0,1,\cdots,7)$ satisfy the CR $[\hat{b}_i, \hat{b}_j] = \delta_{i,j}$. The arrangement for the additional BS's which will produce the transformation (20) is straightforward and can be implemented by following similar arrangement to that presented for the three-qubit discrete Fourier transform (see [6], Figures 4 and 5).

Combination of the transformations (20) and (21) can be represented as

$$\begin{pmatrix} \hat{b}_0 \\ \hat{b}_1 \\ \hat{b}_2 \\ \hat{b}_3 \\ \hat{b}_4 \\ \hat{b}_5 \\ \hat{b}_6 \\ \hat{b}_7 \end{pmatrix} = G \begin{pmatrix} \hat{a}_0 \\ \hat{a}_1 \\ \hat{a}_2 \\ \hat{a}_3 \\ \hat{a}_4 \\ \hat{a}_5 \\ \hat{a}_6 \\ \hat{a}_7 \end{pmatrix}, \qquad (21)$$

where $G$ is a unitary matrix of dimension $8 \times 8$. One should notice from the above analysis that each of the $\hat{b}_i$ $(i = 0,1,\cdots,7)$ is a function of *all* the $\hat{a}_i$ $(i = 0,1,\cdots,7)$.

A *unitary matrix* $G$ of dimension $8 \times 8$ can be represented by the block diagram

$$G = \begin{pmatrix} A & B \\ C & E \end{pmatrix} \quad , \tag{22}$$

where $A$, $B$, $C$, $D$ are $4 \times 4$ matrices which would lead to the relation:

$$GG^\dagger = \begin{pmatrix} A & B \\ C & E \end{pmatrix} \begin{pmatrix} A^\dagger & C^\dagger \\ B^\dagger & E^\dagger \end{pmatrix} = \begin{pmatrix} I & 0 \\ 0 & I \end{pmatrix} \quad , \tag{23}$$

under the conditions:

$$AA^\dagger + BB^\dagger = I \quad , \tag{24}$$

$$CC^\dagger + EE^\dagger = I \quad , \tag{25}$$

$$AC^\dagger + BE^\dagger = 0 \quad , \tag{26}$$

$$CA^\dagger + EB^\dagger = 0 \quad . \tag{27}$$

Since we have used only unitary transformation for getting the final unitary matrix $G$ of dimension $8 \times 8$, relations (22-27) are satisfied for our system. In Eqs. (23-27) $I$ and 0 represent the $4 \times 4$ unit and zero matrix, respectively. In order to implement the quantum gates one should take care only for getting the upper left block $A$ of $G$ of dimension $4 \times 4$ which will correspond to the non-unitary transformation $A_{CNOT}$ of (12) or $A_{SWAP}$ of (14), up to normalization constants. All other matrix elements of the $8 \times 8$ unitary transformation are fixed by the experimental set up which follow the transformations (19) and (20).

We can simplify our equations by using special relations valid in our system. The input photons are inserted only in the input ports 0,1,2, and 3 while only the vacuum is entering in the input ports 4,5,6, and 7. At the end of the multiport transformations the photons can exit in the 8'th output ports. Therefore it is enough to implement the transformations operating only the input operators $\hat{a}_0$, $\hat{a}_1$, $\hat{a}_2$, $\hat{a}_3$ where the input operators $\hat{a}_4$, $\hat{a}_5$, $\hat{a}_6$, $\hat{a}_7$ should be used only for satisfying the CR but they don't affect the photon numbers in the outputs. Following this idea the output operators after the third stage including the 12 BS's can be written as

$$\hat{b}_0 = \hat{\alpha}\cos\theta_9 + \hat{\beta}\sin\theta_9 + \cdots \;;$$
$$\hat{\alpha} = \hat{a}_0 \cos\theta_5 \cos\theta_1 + \hat{a}_2 \sin\theta_5 \cos\theta_3 \;; \quad \hat{\beta} = \hat{a}_1 \cos\theta_6 \cos\theta_2 + \hat{a}_3 \sin\theta_6 \cos\theta_4 \;;$$
$$\hat{b}_1 = \hat{\gamma}\cos\theta_{10} + \hat{\delta}\sin\theta_{10} + \cdots \;;$$
$$\hat{\gamma} = \hat{a}_2 \cos\theta_5 \cos\theta_3 - \hat{a}_0 \sin\theta_5 \cos\theta_1 \;; \quad \hat{\delta} = \hat{a}_3 \cos\theta_6 \cos\theta_4 - \hat{a}_1 \sin\theta_6 \cos\theta_2 \;;$$
$$\hat{b}_2 = \hat{\xi}\cos\theta_{11} + \hat{\eta}\sin\theta_{11} + \cdots \;;$$
$$\hat{\xi} = -\hat{a}_0 \cos\theta_7 \sin\theta_1 - \hat{a}_2 \sin\theta_7 \sin\theta_3 \;; \quad \hat{\eta} = -\hat{a}_1 \cos\theta_8 \sin\theta_2 - \hat{a}_3 \sin\theta_8 \sin\theta_4 \;;$$
$$\hat{b}_3 = \hat{\psi}\cos\theta_{12} + \hat{\phi}\sin\theta_{12} + \cdots \;;$$
$$\hat{\psi} = \hat{a}_0 \sin\theta_7 \sin\theta_1 - \hat{a}_2 \cos\theta_7 \sin\theta_3 \;; \quad \hat{\phi} = \hat{a}_1 \sin\theta_8 \sin\theta_2 - \hat{a}_3 \cos\theta_8 \sin\theta_4 \;;$$
$$\hat{b}_4 = \hat{\beta}\cos\theta_9 - \hat{\alpha}\sin\theta_9 + \cdots \;;$$
$$\hat{b}_5 = \hat{\delta}\cos\theta_{10} - \hat{\gamma}\sin\theta_{10} + \cdots \;;$$
$$\hat{b}_6 = \hat{\eta}\cos\theta_{11} - \hat{\xi}\sin\theta_{11} + \cdots \;;$$
$$\hat{b}_7 = \hat{\phi}\cos\theta_{12} - \hat{\psi}\sin\theta_{12} + \cdots \;.$$

(28)

In Eqs. (28) we have omitted the operators $\hat{a}_4$, $\hat{a}_5$, $\hat{a}_6$, $\hat{a}_7$ which do not affect the photon number outputs but are referred by the notation '$\cdots$' as they are needed only for satisfying the CR. In order to implement the quantum gates we exclude by postselection all the cases in which one or two photons exit through the output ports $\hat{b}_4$, $\hat{b}_5$, $\hat{b}_6$, $\hat{b}_7$.

According to (28) the output operators $\hat{b}_0$, $\hat{b}_1$, $\hat{b}_2$, $\hat{b}_3$ can be given as functions of the operators $\hat{a}_0$, $\hat{a}_1$, $\hat{a}_2$, $\hat{a}_3$ as:

$$\hat{b}_0 = \hat{a}_0 \cos\theta_9 \cos\theta_5 \cos\theta_1 + \hat{a}_1 \sin\theta_9 \cos\theta_6 \cos\theta_2 + \hat{a}_2 \cos\theta_9 \sin\theta_5 \cos\theta_3 + \hat{a}_3 \sin\theta_9 \sin\theta_6 \cos\theta_4 + \cdots;$$
$$\hat{b}_1 = -\hat{a}_0 \sin\theta_5 \cos\theta_1 \cos\theta_{10} - \hat{a}_1 \sin\theta_6 \cos\theta_2 \sin\theta_{10} + \hat{a}_2 \cos\theta_5 \cos\theta_3 \cos\theta_{10} + \hat{a}_3 \cos\theta_6 \cos\theta_4 \sin\theta_{10} + \cdots;$$
$$\hat{b}_2 = -\hat{a}_0 \cos\theta_7 \sin\theta_1 \cos\theta_{11} - \hat{a}_1 \cos\theta_8 \sin\theta_2 \sin\theta_{11} - \hat{a}_2 \sin\theta_7 \sin\theta_3 \cos\theta_{11} - \hat{a}_3 \sin\theta_8 \sin\theta_4 \sin\theta_{11} + \cdots;$$
$$\hat{b}_3 = \hat{a}_0 \sin\theta_7 \sin\theta_1 \cos\theta_{12} + \hat{a}_1 \sin\theta_8 \sin\theta_2 \sin\theta_{12} - \hat{a}_2 \cos\theta_7 \sin\theta_3 \cos\theta_{12} - \hat{a}_3 \cos\theta_8 \sin\theta_4 \sin\theta_{12} + \cdots.$$

(29)

The matrix $A_{CNOT}$ of (12) implements (up to normalization constant) the transformation

$$\sqrt{8}\begin{pmatrix}\hat{b}_0\\\hat{b}_1\\\hat{b}_2\\\hat{b}_3\end{pmatrix}=\begin{pmatrix}\hat{a}_0-\hat{a}_1+\hat{a}_2+\hat{a}_3\\-\hat{a}_0+\hat{a}_1+\hat{a}_2+\hat{a}_3\\\hat{a}_0-\hat{a}_1+\hat{a}_2+\hat{a}_3\\-\hat{a}_0+\hat{a}_1+\hat{a}_2+\hat{a}_3\end{pmatrix}. \quad (30)$$

Therefore the *CNOT* gate will be simulated if the transformation (30) will be equal to the transformation (29) (excluding the terms referred by the notation $\cdots$ ) which corresponds to the upper left block of $G$.

The matrix $A_{SWAP}$ of (14) implements (up to normalization constant) the transformation

$$\sqrt{8}\begin{pmatrix}\hat{b}_0\\\hat{b}_1\\\hat{b}_2\\\hat{b}_3\end{pmatrix}=\begin{pmatrix}\hat{a}_0-\hat{a}_1+\hat{a}_2+\hat{a}_3\\-\hat{a}_0+\hat{a}_1+\hat{a}_2+\hat{a}_3\\-\hat{a}_0+\hat{a}_1-\hat{a}_2+\hat{a}_3\\\hat{a}_0-\hat{a}_1-\hat{a}_2+\hat{a}_3\end{pmatrix}. \quad (31)$$

Therefore the *SWAP* gate will be simulated if the transformation (31) will be equal to the transformation (29) (excluding the terms referred by the notation '$\cdots$' ) which corresponds to the upper left block of $G$.

We find that for simulating the *CNOT* and *SWAP* gates we can use 50:50 BS's which will satisfy the conditions

$$\cos(\theta_i)=|\sin(\theta_i)|=\frac{1}{\sqrt{2}}, \quad (32)$$

so that only the signs of $\sin(\theta_i)$ are chosen to get agreement between (29) and (30) for simulating the *CNOT* gate, and between (29) and (31) for simulating the *SWAP* gate.

The transformation (29) is equivalent to the $A_{CNOT}$ transformation of (30) under the conditions:

$$\sin(\theta_1) = \sin(\theta_3) = \sin(\theta_6) = \sin(\theta_9) = \frac{-1}{\sqrt{2}} \; ;$$

$$\sin(\theta_5) = \sin(\theta_7) = \sin(\theta_{10}) = \frac{1}{\sqrt{2}} \; ;$$

$$\sin(\theta_2)\sin(\theta_{11}) = \frac{1}{2} \; ; \quad \sin(\theta_4)\sin(\theta_{12}) = \frac{-1}{2} \; ; \tag{33}$$

$$\sin(\theta_4)\sin(\theta_8)\sin(\theta_{11}) = \frac{-1}{\sqrt{8}} \; ; \quad \sin(\theta_2)\sin(\theta_8)\sin(\theta_{12}) = \frac{1}{\sqrt{8}}$$

The transformation (29) is equivalent to the $A_{SWAP}$ transformation of (31) under the conditions:

$$\sin(\theta_6) = \sin(\theta_9) = \frac{-1}{\sqrt{2}} \; ;$$

$$\sin(\theta_1) = \sin(\theta_3) = \sin(\theta_5) = \sin(\theta_7) = \sin(\theta_{10}) = \frac{1}{\sqrt{2}} \; ;$$

$$\sin(\theta_4)\sin(\theta_{12}) = \sin(\theta_2)\sin(\theta_{11}) = \frac{-1}{2} \; ; \tag{34}$$

$$\sin(\theta_4)\sin(\theta_8)\sin(\theta_{11}) = \sin(\theta_2)\sin(\theta_8)\sin(\theta_{12}) = \frac{-1}{\sqrt{8}}$$

The probability to get one photon into the outports $\hat{b}_0, \hat{b}_1, \hat{b}_2, \hat{b}_3$ is decreasing due to the postselection by a factor 2. The probability to get the two photons into the output ports $\hat{b}_0, \hat{b}_1, \hat{b}_2, \hat{b}_3$ is reduced by a factor 4 so that only 25% of the experiments are taken into account. A fundamental issue in using postselection is that we should not make any measurements on the output ports $\hat{b}_0, \hat{b}_1, \hat{b}_2, \hat{b}_3$ but only verify that one or two photons are not exiting into the output ports $\hat{b}_4, \hat{b}_5, \hat{b}_6, \hat{b}_7$.

## 4. DISCUSSION, SUMMARY AND CONCLUSION

In the present work we have used a special definition for the qubit states $|0\rangle$ and $|1\rangle$. A single photon entering into one input port of a BS is considered as the $|0\rangle$ state while a single photon entering into the second input port of the BS is considered as the $|1\rangle$ state. Such definition has been used in multiport BS's configurations for certain applications in quantum computation [2-6]. For applying such definition for two-qubit states we need to use two BS's which include four states $|00\rangle, |01\rangle, |10\rangle, |11\rangle$ where the first and second number in the ket state denote the input ports into which two single photons are entering in the first and second BS, respectively. Instead of using the 4 dimensional vectors of (6) we are using for interference applications the 4 dimensional states given by (11) as superposition states. Such states are realized by two single photons, each entering in a different BS

Multiport BS's configurations cannot realize directly the quantum gates which are based on *controlled* operations. In the present study we have shown, however, that by adding post-selection [16-43] to the unitary multiport BS's transformations we can simulate the quantum gates. We have analyzed the use of this method for simulating the *CNOT* and the *SWAP* gates, but similar methods can be used for other gates. We find the interesting result that *CNOT* (up to certain relative phase) and the SWAP gate, can be *simulated mathematically* by the *non-unitary* matrix transformations $A_{CNOT}$ of Eqs.(12) and $A_{SWAP}$ of Eqs.(14), respectively.

In order to simulate the quantum gates by extended unitary matrix with post selection we assume to have initially 8 input operators $\hat{a}_0, \hat{a}_1, \hat{a}_2 \ldots \hat{a}_7$ which are mixed by the multiport configurations to produce the output operators $\hat{b}_0, \hat{b}_1, \hat{b}_2 \ldots \hat{b}_7$, which will be related to the input operators by a unitary transformation. For the purpose of simulating the *CNOT* and the *SWAP* gates we can apply for each BS the simple transformation given by (18). The 12 BS's used in this analysis are characterized by the corresponding angles $\theta_1, \theta_2, \theta_3 \cdots \theta_{12}$.

The use of 12 BS's leads to quite complicated transformations represented by Eqs. (19-20). For the present analysis we can use, however, simplifying conditions. The input photons are inserted only in the input ports $0, 1, 2$ and $3$, while only the vacuum is entering in the input ports $4, 5, 6$ and $7$. Therefore it is enough to implement the transformations operating only the input operators $\hat{a}_0, \hat{a}_1, \hat{a}_2, \hat{a}_3$ where the input operators $\hat{a}_4, \hat{a}_5, \hat{a}_6, \hat{a}_7$ should be used only for satisfying the CR but they don't affect the photon numbers in the outputs. Following this idea the output operators after the transformations obtained by the 12 BS's has been written as Eq. (28).

The full unitary transformation performed by the 12 BS's is presented by the matrix $G$ of dimension $8 \times 8$ with its properties given by Eqs. (22-27). For simulating the quantum gates it is, however, enough to equate the transformation (29) (performed by the upper left block of G denoted in (23) as $A$) with the *non-unitary* transformation given by $A_{CNOT}$ of (12) or $A_{SWAP}$ of (14). We find by using such comparisons that for these two gates the conditions (32) are satisfied so that by choosing the signs of $\sin(\theta_i)$, $\theta_i = 1, 2 \cdots, 12$ it is enough to get the non-unitary transformation. For simulating the *CNOT* gate we use the conditions (33) while for the simulating the SWAP gate we get the conditions (34).

In order to simulate the quantum gates we exclude by postselection all the cases in which one or two photons exit through output ports $\hat{b}_4, \hat{b}_5, \hat{b}_6, \hat{b}_7$. So, we find in our analysis that only 25% of the experiments are taken into account. An essential point in quantum computation is that the two-qubit quantum gate should be realized without any measurement on it. Since in the post selection no measurement is done on the output ports $\hat{b}_0$, $\hat{b}_1$, $\hat{b}_2$, $\hat{b}_3$ such requirement is satisfied.

The present analysis is based on the use of controlling multiport transformation with 12 BS's. It might be quite difficult to realize the controlling of such system but once such physical system is established many quantum computation effects can be realized. In a previous work it has been shown how to implement in such system three-qubits discrete Fourier transform [6]. In the present work it has been shown that such system can simulate the *CNOT* and the SWAP gates by the use of postselection. By generalizing the BS's transformations (18) many other quantum gates and quantum effects can be simulated.